\documentclass[twocolumn,aps,superscriptaddress,showpacs,nofootinbib,floatfix]{revtex4-1}
\usepackage{epsfig,bm,feynmf}
\usepackage{graphics}
\usepackage{amsmath}
\usepackage[eps]{pstricks}
\usepackage{physics}
\usepackage{slashed}
\usepackage{dsfont}
\usepackage{multirow}
\usepackage{hyperref}
\usepackage{subfigure}
\usepackage[normalem]{ulem}  

\renewcommand\sout{\bgroup \color{red} \ULdepth=-.5ex \ULset}

\newcommand{\als}{\alpha_{s}}

\newcommand{\gm}[1][\mu]{\gamma_#1}
\newcommand{\ld}[1][a]{\lambda^#1}
\newcommand{\glucon}{\ev{\frac{\alpha_{s}}{\pi}G^{2}}}

\begin{document}


\title{$\phi$ meson properties in nuclear matter from QCD sum rules with chirally separated four-quark condensates }


\author{Jisu Kim}%
\email{fermion0514@yonsei.ac.kr}
\affiliation{Department of Physics and Institute of Physics and Applied Physics, Yonsei University, Seoul 03722, Korea}

\author{Philipp Gubler}%
\email{gubler@post.j-parc.jp}
\affiliation{Advanced Science Research Center, Japan Atomic Energy Agency, Tokai, Ibaraki 319-1195, Japan}

\author{Su Houng Lee}%
\email{suhoung@yonsei.ac.kr}
\affiliation{Department of Physics and Institute of Physics and Applied Physics, Yonsei University, Seoul 03722, Korea}


\begin{abstract}
The modification of the $\phi$ meson spectrum in nuclear matter is studied in an updated QCD sum rule analysis, taking into account recent 
improvements in properly treating the chiral invariant and breaking components of four-quark condensates. Allowing both mass and decay width to change at finite density, the QCD sum rule analysis determines 
certain combinations of changes for these parameters that satisfy the sum rules equally well.  A comprehensive 
error analysis, including uncertainties related to the behavior of various condensates at linear order in density, the employed renormalization scale and perturbative corrections of the Wilson coefficients, is used to compute the allowed ranges of these parameter combinations. 
We find that the $\phi$ meson mass shift in nuclear matter is especially sensitive to the strange sigma term $\sigma_{sN}$, which determines the decrease of the strange quark condensate in nuclear matter. Specifically, we obtain a linear relation between the width $\Gamma_{\phi}$ and mass shift $\Delta m_{\phi}$ given as $ \Gamma_{\phi} =  a\Delta m_{\phi} + b\sigma_{sN}+c$ 
with $a = (3.947^{+0.139}_{-0.130})$, $b = (0.936^{+0.180}_{-0.177} )$ and $c = -(7.707^{+4.791}_{-5.679}) \mathrm{MeV}$.

\end{abstract}


\maketitle

\section{Introduction}

The emergence of hadron masses larger than a few hundred MeV from quarks lighter than 
10 MeV and gluon dynamics governed by the strong interaction is one of the still not 
fully understood phenomena of QCD \cite{Wilczek:1999be,Wilczek:2012sb}. 
While the effect of spontaneous chiral symmetry breaking \cite{Nambu:1961tp,Nambu:1961fr} is 
widely believed to be at least in part responsible for the hadronic mass generation 
\cite{Hatsuda:1985eb,Brown:1991kk,Hatsuda:1991ez,Leupold:2009kz}, it remains difficult 
to experimentally confirm this theoretical scenario.

One promising experimental strategy that has been pursued during the past decades is to study 
the behavior of hadron masses in extreme conditions such as finite temperature and/or density, 
because chiral symmetry is expected to be at least partially restored in such environments 
(see Ref.\cite{Gubler:2018ctz} for a recent review). If hadron masses are hence really generated even partially by chiral symmetry breaking, they should be modified in hot and/or dense matter. 
Furthermore, chiral symmetry breaking can naturally explain the mass difference between chiral 
partners \cite{Weinberg:1967kj}, so that if chiral symmetry is restored, the mass difference between 
them can be expected to vanish \cite{Song:2018plu,Lee:2019tvt}. 
Based on these motivations, worldwide experiments have indeed been 
performed attempting to measure hadronic mass shifts at finite temperature or density 
\cite{Hayano:2008vn,Metag:2017yuh,Ichikawa:2018woh,Ohnishi:2019cif,Salabura:2020tou}. 

In this work, we focus on the finite density behavior of the $\phi$ meson, which has recently 
attracted much theoretical 
\cite{Gubler:2014pta,Gubler:2015yna,Gubler:2016itj,Cabrera:2016rnc,Cabrera:2017agk,Cobos-Martinez:2017vtr,Cobos-Martinez:2017woo,Kim:2017nyg,Kim:2019ybi} 
and experimental 
\cite{Ishikawa:2004id,KEK-PS-E325:2005wbm,E325:2006ioe,CLAS:2010pxs,Polyanskiy:2010tj,Hartmann:2012ia,Hartmann:2012ia,HADES:2018qkj,Ashikaga:2019jpc,ALICE:2021cpv} 
attention. 
On the theoretical side, following the original works of Refs.\cite{Hatsuda:1991ez,Lee:1997zta}, 
the QCD sum rule analysis of Ref.\cite{Gubler:2014pta} demonstrated that a clear relation exists between 
the $\phi$ meson mass shift in nuclear matter and the strange sigma term $\sigma_{sN}$, which 
governs the decrease of the strange quark condensate within the linear density approximation, 
$\langle \overline{s}s \rangle_{\rho} \simeq \langle \overline{s}s \rangle_{0} + \frac{\sigma_{sN}}{m_s} \rho$. 
Furthermore, Ref.\cite{Kim:2019ybi} showed that the longitudinal and transverse modes of the 
$\phi$ meson, which at finite density generally become independent due to the broken Lorentz 
symmetry, have different non-trivial dispersion relations, which could be manifested in 
distinct peaks in the respective dilepton spectrum for non-zero spatial momentum with 
respect to the nuclear matter rest frame. Moreover, there is a multitude of recent works studying the 
$\phi$ meson spectral function at finite density using hadronic effective theories  \cite{Gubler:2015yna,Gubler:2016itj,Cabrera:2016rnc,Cabrera:2017agk,Cobos-Martinez:2017vtr,Cobos-Martinez:2017woo}. 

Experimentally, no consensus has yet been reached on how the $\phi$ meson behaves in  nuclear matter. 
While the KEK E325 experiment measured for the $\phi$ meson peak both a negative mass shift and 
broadening at normal nuclear matter 
density \cite{KEK-PS-E325:2005wbm}, no mass shift was seen at HADES \cite{HADES:2018qkj}, which 
rather obtained data suggesting only broadening effects, along the line of earlier results reported 
in Refs.\cite{Ishikawa:2004id,CLAS:2010pxs,Polyanskiy:2010tj}. On the other hand, the recent ALICE measurement 
of the $\phi N$ correlation 
function in Ref.\cite{ALICE:2021cpv} led to a determination of the $\phi N$ channel scattering length with a large 
real part corresponding to an attractive interaction (hence suggesting a negative mass shift) and 
a small imaginary part (suggesting weak broadening effects). Interestingly, the absolute value 
of the thus obtained scattering length is much larger than what was measured in earlier 
$\phi$ meson photoproduction experiments \cite{Chang:2007fc,Strakovsky:2020uqs}. 
It is hoped that the situation will be clarified at the ongoing J-PARC E16 experiment, where 
pA reactions with various target nuclei will be probed to measure the finite density dilepton 
spectrum in the $\phi$ meson mass region \cite{Ashikaga:2019jpc}. There is also a proposal 
to measure at J-PARC not only the dilepton, but also the $K^{+}K^{-}$ spectrum from the same pA 
reactions, which would serve not only as an important consistency check, but also provide 
opportunities for new measurements thanks to the much higher statistics and to the larger branching 
ratio of this channel with different angular dependencies of the outgoing particles, which 
might allow the separation of the longitudinal and transverse modes \cite{P88_JPARC}.
 
For the theoretical study in this paper we make use of the QCD sum rule method, including the 
following two major improvements compared to previous works. First, we take into account the possibility 
of broadening in the phenomenological parametrization of the spectral function, 
following similar approaches in Refs.\cite{Leupold:1997dg,Morita:2007pt}. 
As discussed already in these previous works, the broadening 
effect has to be introduced as a new and independent parameter in the sum rule analysis. It is then 
no longer possible to uniquely determine both mass shift and width of the $\phi$ meson from the sum rules, 
but strong correlations between them. Second, we consider the chiral properties of the four-quark condensates 
that appear in the $\phi$ meson sum rules by separating them into chiral symmetric (e.g. invariant) and 
chiral breaking parts and treating the two contributions independently. This allows us to 
study the role of chiral symmetry restoration not only from the conventional chiral condensate of mass dimension 3, 
but also the dimension 6 four-quark condensates \cite{Kim:2020zae}. Moreover, it becomes possible to go beyond the usually 
applied vacuum saturation approximation. 
Two of the present authors have already applied this advanced treatment of the four-quark condensates to the 
study the finite density behavior of light vector and axial-vector mesons and nucleon and delta resonances \cite{Kim:2021xyp} and 
have for instance found that the masses of the chiral partners $\rho$ and $a_1$ become 
degenerate with a mass of about 550 to 600 MeV in the vacuum with restored chiral symmetry \cite{Kim:2020zae}. 

This paper is organized as follows. In Section\,\ref{sec:separation}, we recapitulate our treatment of the four-quark 
condensates, paying attention to the specific condensates needed for the $\phi$ meson sum rule. This is followed by 
a general description the QCD sum rule approach in Section\,\ref{sec:qcs_sum_rules}. The results of this work are summarized in Section\,\ref{sec:results}, starting with 
a discussion of the vacuum $\phi$ meson sum rule in Subsection\,\ref{sec:vacuum}, which provides the baseline of the later 
finite density analysis, given in Subsection\,\ref{sec:nuclear_matter}. The main features and physical consequences of the obtained findings are discussed in Section\,\ref{sec:discussion}, after which the paper is concluded in Section\,\ref{sec:conclusion}. 

\section{Separation of four-quark condensates \label{sec:separation}}
In this section, we summarize how to separate the four-quark condensates into chiral symmetric and breaking parts, which was already used in Ref.\,\cite{Kim:2020zae} to study the $\rho$ and $a_1$ meson channels. 
The  quark propagator $S_q(0,x)$ appearing inside an expectation value can always be decomposed into chiral symmetric and breaking parts by adding and subtracting its chiral partner, which are also called the chiral even $(S)$  and odd $(B)$  components, respectively.
\begin{eqnarray}
S_q(0,x) &=&  \bigg( S_q^B(0,x)+ S_q^S(0,x) \bigg),  \label{prop0} \\
S_q^B(0,x) &=& \frac{1}{2} \bigg( S_q(0,x)-i \gamma^5 S_q(0,x)i\gamma^5 \bigg),  \\
S_q^S(0,x) &=& \frac{1}{2} \bigg( S_q(0,x)+i \gamma^5 S_q(0,x)i\gamma^5 \bigg). 
\label{porp}
\end{eqnarray}
For the dimension 3 two-quark condensate only the chiral symmetry breaking part contributes due to the trace, 
\begin{equation} 
\begin{split} 
\ev{\bar{q} q} &= - \lim_{x \rightarrow 0}  \langle \Tr [S_q^B(0,x)]
\rangle \\
& = -\pi \rho(0),
\label{bc-rel}
\end{split} 
\end{equation}
where $\rho(0)$ the density of  zero modes as derived by Casher and Banks \cite{BC}. 

A four-quark condensate typically has the following general form, which can be written in terms of disconnected $(dis)$ and connected $(con)$ pieces of the quark propagator denoted by the respective subscripts,  
\begin{equation} 
\begin{split} 
\ev{(\bar{q} \Gamma q)(\bar{q} \Gamma q)} &= \langle \Tr[S_{q}^i\Gamma]\Tr[S_{q}^i\Gamma] \rangle_{dis} - \langle \Tr[\Gamma S_{q}^i \Gamma S_{q}^i] \rangle_{con},
\label{4-q_sep}
\end{split} 
\end{equation}
where $\Gamma$ is an arbitrary matrix that can contain Dirac, color and/or flavor components. The summation in $i=B,S$ does not include cross terms as the Dirac matrix in $\Gamma$ will 
only allow one type to contribute. 
As shown in Eq.~\eqref{bc-rel}, $S_q^B$ is proportional to density of zero modes so that a general four-quark 
condensate can be expressed as,
\begin{equation} 
\begin{split} 
\ev{(\bar{q} \Gamma q)(\bar{q} \Gamma q)} = \ev{(\bar{q} \Gamma q)(\bar{q} \Gamma q)}_B+\ev{(\bar{q} \Gamma q)(\bar{q} \Gamma q)}_S,
\end{split} 
\end{equation}
where the subscript $B$ and $S$ represent the chiral symmetry breaking and symmetric parts, respectively, and the former is proportional to $\rho^{2}(0)$.

Let us now consider the four-quark condensates appearing in the $\phi$ meson sum rule which are the following three types,
\begin{equation} 
\begin{split} 
\ev{(\bar{s} \gm \ld s)^{2}},\;\ev{(\bar{s} \gm \gm[5] \ld s)^{2}}, \;\ev{(\bar{s} \gm \ld s)(\bar{u} \gm \ld u)}.
\end{split} 
\end{equation}
For the condensate $\ev{(\bar{s} \gm \ld s)(\bar{u} \gm \ld u)}$, as is clear from its flavor structure, only the disconnected term contributes.  Since the symmetry breaking  quark propagator vanishes due to the trace $\Tr[\gamma_{\mu} S_q^B]=0$, it is a chiral symmetric condensate, 
\begin{equation} 
\begin{split} 
\ev{(\bar{s} \gm \ld s)(\bar{u} \gm \ld u)} = \ev{(\bar{s} \gm \ld s)(\bar{u} \gm \ld u)}_{dis,S}.
\end{split} 
\end{equation}
The other two condensates have contributions from both the chiral symmetry breaking and symmetric parts. One first notes that, both have disconnected  and connected contributions.
For the connected piece, using the property 
\begin{eqnarray}
i \gamma^5 S_s^B i \gamma^5 =-S_s^B,
\end{eqnarray}
one can write the breaking part in the two condensates as a single operator with an overall sign depending 
on its Dirac structure. In summary, one finds
\begin{equation} 
\begin{split} 
\ev{(\bar{s} \gm \ld s)^{2}} =&\; \ev{(\bar{s} \gm \ld s)^{2}}_{dis,S} +\ev{(\bar{s} \gm \ld s)^{2}}_{con,S}\\
&+ \ev{(\bar{s} \gm \ld s)^{2}}_{con,B},\\
\ev{(\bar{s} \gm \gm[5] \ld s)^{2}} =&\; \ev{(\bar{s} \gm \gm[5] \ld s)^{2}}_{dis,S} + \ev{(\bar{s} \gm \ld s)^{2}}_{con,S}\\
&-\ev{(\bar{s} \gm \ld s)^{2}}_{con,B}.
\end{split} 
\end{equation}

In the following analysis, we will first use the experimental values of the mass and width of the meson of interest 
to estimate the value of the  four-quark condensates  appearing in its vacuum sum rule. 
As discussed, the four-quark condensates will be composed of the chiral symmetric and breaking parts.  We will then use the vacuum saturation hypothesis to estimate the values of the four-quark operators.  Within this approximation, the symmetric operators will vanish, while the breaking operators will become proportional to the square of the  quark condensate, $\ev{\bar{q} q}^2$.
For convenience, we introduce an auxiliary parameter $\kappa$ defined as their ratio. 
\begin{equation} 
\begin{split} 
\kappa = \frac{\mathcal{M}}{\mathcal{M}_{v.s.}},
\end{split} 
\end{equation}
where $\mathcal{M}$ is the total magnitude of the four-quark condensates extracted from our sum rule method and $\mathcal{M}_{v.s.}$ is the total magnitude expected from the vacuum saturation hypothesis. $\kappa=1$ corresponds to the case of using the vacuum saturation of the four-quark condensates.

For the sum rules of the $(\rho, a_1)$ mesons that form chiral partners, the four-quark condensates contributing to the respective OPE can be shown to have the same contribution from the chirally symmetric part but a breaking part contributing with different coefficients.  Therefore, both contributions can be determined  uniquely. 

It should be noted that not all vector mesons ($J^P=1^-$) have chiral partners with $J^P=1^+$. This is so because the vector meson octet mixes with the singlet.  For example, an ideally mixed $\omega$ meson is invariant under SU(2) chiral rotation, while the $\rho$ mixes with the $a_1$\cite{Gubler:2016djf}.  
This implies that the number of independent combinations of four-quark operators appearing in the $(\omega,f_1^{\bar{q}q})$ sum rules is more than two so that determining them from the two parity sum rules is no longer possible. On the other hand,  as 
the difference between $(\rho, a_1)$  and $(\omega,f_1^{q\bar{q}})$ sum rules are due to the disconnected diagrams, one can estimate these matrix elements from comparing the  respective phenomenology and thus obtain the values of all  combinations of chiral  breaking and symmetric operators appearing in the sum rules for the latter pair.  Furthermore, assuming the contributions of disconnected diagrams to be flavor independent, one can estimate the values of chiral symmetry  breaking and symmetric operators appearing in the $(\phi,f_1^{s\bar{s}})$ as well as all other spin 1 sum rules \cite{Kim:2021xyp}.

\section{QCD sum rules analysis \label{sec:qcs_sum_rules}}
We start with the two-point correlation function of the interpolating currents, which couple the meson states to be investigated in this work,
\begin{equation} 
\begin{split} 
\Pi_{\mu \nu}(q) &= i \int d^{4} e^{iqx} \ev{T[j_{\mu}(x)j_{\nu}(0)]}.
\end{split} 
\label{eq:correlator}
\end{equation}
For the $\phi$ and $f_{1}^{s\bar{s}}$ meson, the currents $j_{\mu} = \bar{s}\gm s$ and $j_{\mu}= \bar{s} \gm \gm[5] s$ are used, respectively, where only the transverse component is taken for the axial vector current~\cite{Gubler:2016djf}. 
If the bracket $\ev{}$ has the explicit subscript $\rho$, it stands for the expectation value with respect to the 
ground state of nuclear matter 
at density $\rho$ and zero temperature. Otherwise, it stands for the vacuum expectation value. In both vacuum and in-medium cases, we only study 
the contracted correlation function defined as
\begin{equation} 
\begin{split} 
\Pi(q^{2}) = \frac{1}{3q^{2}}\Pi^{\mu}_{\mu}(q),
\end{split} 
\end{equation}
which is sufficient, as long as we are interested in the case of the meson at rest with respect to the medium.

The phenomenological side is constructed using the following parametrizations of the spectral  function $\rho(s)$, 
 \begin{equation}
 \begin{split}
 \rho(s) &= \rho^{\text{pole}}(s)	 + \rho^{\text{cont}}(s), \\
\rho^{\mathrm{pole}}(s) &= \frac{1}{\pi}
\frac{f \Gamma \sqrt{s}}{(s - m^2)^2 + s\Gamma^2}\theta(s_0 - s),  \label{pheno_side_pole}
\\
\rho^{\mathrm{cont}}(s) &= \frac{1}{\pi} \theta(s - s_0) \mathrm{Im} \tilde{\Pi}^{\mathrm{pert}}(s),
\end{split}
\end{equation}
where $m$ ($\Gamma$) is the single hadron Breit-Wigner mass (width) and $s_{0}$ is the threshold parameter. The values of $m$ and $\Gamma$ for the $\phi$ ($f_{1}^{s\bar{s}}$) meson employed to construct the pole structure are 1019.45 MeV (1426.3 MeV) and 4.26 MeV (54.5 MeV), respectively. These will be used to determine the values of the four-quark operators appearing in the dimension-6 term in the operator product expansion (OPE) 
of Eq.~(\ref{eq:correlator}). After the Borel transform, the phenomenological side is given by
\begin{equation} 
\begin{split} 
\widehat{\Pi}^{\mathrm{pole,\;cont}}(M^2) = \int_{4m_{\pi}}^{\infty} ds \;e^{-s/M^2} \rho^{\mathrm{pole,\;cont}}(s),
\end{split} 
\end{equation}
where $m_{\pi}$ is the pion mass and $M$ the non-physical Borel mass. 
The corresponding Borelized OPE side is obtained as
\begin{equation} 
\begin{split} 
\widehat{\Pi}^{\mathrm{OPE}}(M^2) =& c_{0} + \frac{c_{2}}{M^{2}} +\frac{c_{4}}{M^{4}} + \frac{c_{6}}{M^{6}}, 
\end{split} 
\end{equation}
where all terms with dimension 8 and higher are neglected. The coefficients $c_n$ for the $\phi$ meson sum rules in vacuum are given as \cite{Gubler:2014pta}
\begin{equation} 
\begin{split} 
c_{0}(0) =&\; \frac{1}{4\pi^{2}} \bigg(1+\frac{\als}{\pi}+\frac{\als^{2}}{\pi^{2}}\\
&\;\times\bigg[\frac{365}{24}-11\zeta(3) - 3\bigg(\frac{11}{12}-\frac{2}{3}\zeta(3)\bigg)\bigg] \bigg),\\
c_{2}(0) =&\; -\frac{3m_{s}^{2}}{2\pi^{2}}\bigg(1+\frac{2}{3}\frac{\als}{\pi}\bigg[ 4-6\log (\frac{M}{\mu})+3\gamma_{E}\bigg] \bigg),\\
c_{4}(0) =&\; \frac{1}{12}\bigg(1+\frac{7}{6}\frac{\als}{\pi} \bigg)\ev{\frac{\als}{\pi}G^{2}} + 2m_{s}\bigg(1+\frac{1}{3}\frac{\als}{\pi}\bigg)\ev{\bar{s}s},\\
c_{6}(0) =&\; -\pi \als \bigg[ \ev{(\bar{s} \gm \gm[5] \ld s)^{2}} \\
&\;\quad\qquad+ \frac{2}{9} \ev{(\bar{s}\gm \ld s)(\sum_{q=u,d,s}(\bar{q}\gm \ld q))}\bigg]\\
&+\frac{m_{s}^{2}}{6}\bigg[\frac{1}{3} \ev{\frac{\als}{\pi}G^{2}} - 8m_{s} \ev{\bar{s}s} \bigg].
\end{split} 
\label{eq:OPE_coeff}
\end{equation}
The corresponding coefficients for the $f_{1}^{s\bar{s}}$ meson channel are \cite{Gubler:2016djf}
\begin{equation} 
\begin{split} 
c_{0}^{A}(0) =&\; c_{0}(0)\\
c_{2}^{A}(0) =&\; c_{2}(0)\\
c_{4}^{A}(0) =&\; \frac{1}{12}\bigg(1+\frac{7}{6}\frac{\als}{\pi} \bigg)\ev{\frac{\als}{\pi}G^{2}} - 2m_{s}\bigg(1+\frac{1}{3}\frac{\als}{\pi}\bigg)\ev{\bar{s}s},\\
c_{6}^{A}(0) =&\; -\pi \als \bigg[ \ev{(\bar{s} \gm  \ld s)^{2}} \\
&\;\quad\qquad+ \frac{2}{9} \ev{(\bar{s}\gm \ld s)(\sum_{q=u,d,s}(\bar{q}\gm \ld q))}\bigg]\\
&+\frac{m_{s}^{2}}{6}\bigg[\frac{1}{3} \ev{\frac{\als}{\pi}G^{2}} + 8m_{s} \ev{\bar{s}s} \bigg].
\end{split} 
\end{equation}
The parameter values used in the numerical analysis, to be discussed below, are given
in Table\,\ref{sr_parameter}.
\begin{center}
\begingroup
\setlength{\tabcolsep}{12pt} 
\renewcommand{\arraystretch}{1.5} 
\begin{table}[htbp]
\begin{tabular}{c  l l} \hline
	\hline
$\mu$ & 1 GeV & 2 GeV\\ \hline
$m_{s}$\cite{ParticleDataGroup:2020ssz} & 125.55 MeV & 93 MeV \\
$\als$\cite{Bethke:2009jm} &0.5 &0.31\\
$\ev{\bar{u}u}$\cite{Aoki:2021kgd} & ($-$0.246 GeV)$^3$& ($-$0.272 GeV)$^3$\\
$\ev{(\als/\pi) G^{2}}$ \cite{Colangelo:2000dp}& 0.012 GeV$^4$& 0.012 GeV$^4$\\
 $\ev{\bar{s}s}/\ev{\bar{u}u}$\cite{Reinders:1984sr} & 0.8 & 0.8\\
$M_{N}$          & 939 MeV & 939 MeV\\
$\rho_{0}$       & 0.17 fm$^{-3}$ & 0.17 fm$^{-3}$\\
$\sigma_{\pi N}$ & 45 MeV & 45 MeV\\
$A_{2}^{s}$      & 0.053 & 0.072\\
$A_{2}^{g}$      & 0.367 & 0.425\\
$A_{4}^{s}$      & 0.00121& 0.00122\\
$X$ & 6.40 MeV & 6.40 MeV\\
	 \hline 
\end{tabular}
\caption{Parameter values used for the in-vacuum and/or -medium numerical analyses, for the two
renormalization scales of $\mu = 1\,$GeV and $\mu = 2\,$GeV.
The quark mass and condensate value are scaled to $\mu = 1\,$GeV from the values at $\mu = 2\,$GeV, using the scale conversion factor 1.35, following Ref.\,\cite{ParticleDataGroup:2020ssz}.
$A_{2}^{s}$, $A_{2}^{g}$ and $A_{4}^{s}$ can be found in \cite{Gubler:2018ctz}. Details of the twist-4 operators contained in $X$ are discussed in \cite{Gubler:2015uza}. }
\label{sr_parameter}
\end{table}
\end{center}
\endgroup
To eliminate the dependence on the residue $f$ in $\rho^{\mathrm{pole}}(s)$ shown in  Eq.~(\ref{pheno_side_pole}), we conduct an analysis using the ratio between $\widehat{\Pi}^{\mathrm{phen./OPE}}(M^{2})$ and its derivative with respect to $-1/M^{2}$. 
Specifically, we will use the following form throughout this paper,
\begin{equation} 
\begin{split} 
\frac{\widehat{\Pi}^{\mathrm{pole}}(M^{2})'}{\widehat{\Pi}^{\mathrm{pole}}(M^{2})}=\frac{\widehat{\Pi}^{\mathrm{OPE}}(M^{2})'-\widehat{\Pi}^{\mathrm{cont}}(M^{2})'}{\widehat{\Pi}^{\mathrm{OPE}}(M^{2})-\widehat{\Pi}^{\mathrm{cont}}(M^{2})},
\label{phen_der}
\end{split} 
\end{equation}
where the primes denote the derivative with respect to $-1/M^{2}$.
In vacuum, the experimental values of the mass and decay width are used to evaluate the parameter $\kappa$. In nuclear matter, Eq.~(\ref{phen_der}) is used to extract the in-medium mass 
of the $\phi$ meson for varying decay widths.

To make sure that the various approximations applied to derive the sum rules are valid, one usually defines an appropriate range of the Borel mass $M$ (called the ``Borel window"), which we discuss below. 
On the phenomenological side, one has to make sure that the contribution from the pole $\rho^{\mathrm{pole}}(s)$ dominates over that from the continuum $\rho^{\mathrm{cont}}(s)$. The corresponding constraint is 
\begin{equation} 
\begin{split} 
\frac{\int_{s_0}^{\infty}ds\; e^{-s/M^{2}}\rho^{\mathrm{cont}}(s)}{\int_{0}^{\infty}ds\; e^{-s/M^{2}}\rho^{\mathrm{pole}}(s)} < 0.5,
\label{bm_max}
\end{split} 
\end{equation}
which determines the upper boundary $M_{\mathrm{max}}$ of the Borel window. 
On the other hand, to ensure a reasonable convergence of the OPE, we require the contribution of the condensate terms to be smaller than that of the perturbative term. Specifically,
\begin{equation} 
\begin{split} 
\frac{\widehat{\Pi}^{\mathrm{OPE}}_{\mathrm{cond\; terms}}(M^{2})}{\widehat{\Pi}^{\mathrm{OPE}}_{\mathrm{pert\; term}}(M^{2})} < 0.1,
\label{bm_min}
\end{split} 
\end{equation}
where $\widehat{\Pi}^{\mathrm{OPE}}_{\mathrm{cond\; terms}}(M^{2})$ and $\widehat{\Pi}^{\mathrm{OPE}}_{\mathrm{pert\; term}}(M^{2})$ are the sum of the condensate terms considered and the perturbative term on OPE side, respectively. This condition restricts the Borel mass $M$ to be larger than $M_{\mathrm{min}}$. 

The threshold parameter $s_{0}$ is chosen so that the minimum mass in the Borel curve for mass is located at the center of the Borel window. The mass and kappa values are taken at the extremum point.

\section{Results \label{sec:results}}
\subsection{Vacuum Analysis \label{sec:vacuum}}

To apply the method proposed in Ref.~\cite{Kim:2021xyp} to the $\phi$ and $f_{1}^{s\bar{s}}$(1420) meson channels, we need to extend it from $N_{f}=2$ to the $N_{f} =3$ case. The fundamental difference between the two is whether or not the strange quark condensates are considered sensitive to the change of the zero mode density. Similar to the ($\omega(782)$, $f_{1}^{q\bar{q}}(1285)$) pair, they are not exact chiral partners, but parity partners. They therefore have different chirally symmetric dimension-6 (four-quark) condensates which contribute to their mass difference in the chiral symmetry restored vacuum. They read
\begin{equation} 
\begin{split}
\mathcal{M}_{\phi} =&\; \frac{7}{9} \ev{B_{ss}}_{B}+\ev{(\bar{s}\gm \gm[5] \ld s)^{2}}_{dis,S} +\ev{S}_{S},\\
\mathcal{M}_{f_{1}^{s\bar{s}}} =&\; -\frac{11}{9} \ev{B_{ss}}_{B} +\ev{(\bar{s}\gm  \ld s)^{2}}_{dis,S}+\ev{S}_{S},
\label{phi_f1_4q}
\end{split} 
\end{equation}
where 
\begin{equation} 
\begin{split} 
\ev{(\bar{s} \gm \ld s)^{2}} =&\; \ev{(\bar{s} \gm \ld s)^{2}}_{dis,S} + \ev{(\bar{s} \gm  \ld s)^{2}}_{con,S}\\
& -\ev{(\bar{s} \gm \gm[5] \ld s)^{2}}_{con,B},\\
\ev{(\bar{s} \gm \gm[5] \ld s)^{2}} =&\; \ev{(\bar{s} \gm \gm[5] \ld s)^{2}}_{dis,S} + \ev{(\bar{s} \gm \ld s)^{2}}_{con,S}\\
& +\ev{(\bar{s} \gm \gm[5] \ld s)^{2}}_{con,B},\\
\ev{B_{ss}}_{B} \equiv &\; \ev{(\bar{s} \gm \gm[5] \ld s)^{2}}_{con,B},\\
\ev{S}_{S} \equiv&\; \frac{11}{9}\ev{(\bar{s}\gm \ld s)^{2}}_{con,S} \\
&+\frac{2}{9} \ev{(\bar{s}\gm \ld s)^{2}}_{dis,S}\\
& +\frac{2}{9} \ev{(\bar{s} \gm \ld s)(\bar{l} \gm \ld l)}_{S},
\end{split} 
\end{equation}
and the subscripts are defined as before. Employing the auxiliary parameter $\kappa$, Eq.~\eqref{phi_f1_4q} can be rewritten as
\begin{equation} 
\begin{split} 
\mathcal{M}_{\phi} =&\; \kappa_{\phi}\frac{112}{81}\ev{\bar{s}s}^{2},\\
\mathcal{M}_{f_{1}^{s\bar{s}}}=&\; -\kappa_{f_{1}^{s\bar{s}}}\frac{176}{81}\ev{\bar{s}s}^{2}.
\label{phi_f1_kp}
\end{split} 
\end{equation}
The numerical factors appearing above after the $\kappa$'s are obtained from the vacuum saturation hypothesis. As mentioned in Section\,\ref{sec:separation}, $\kappa_{\phi}$=1, which corresponds to the exact vacuum saturation, is not likely to be a realistic estimate, due to the ignored intermediate states, particularly those involving kaons. 
Combining Eqs.~\eqref{phi_f1_4q} and \eqref{phi_f1_kp}, we have

\begin{equation} 
\begin{split} 
 \kappa_{\phi}\frac{112}{81}\ev{\bar{s}s}^{2} =&\; \frac{7}{9} \ev{B_{ss}}_{B}+\ev{(\bar{s}\gm \gm[5] \ld s)^{2}}_{dis,S} +\ev{S}_{S},\\
 -\kappa_{f_{1}^{s\bar{s}}}\frac{176}{81}\ev{\bar{s}s}^{2} =& -\frac{11}{9} \ev{B_{ss}}_{B} +\ev{(\bar{s}\gm  \ld s)^{2}}_{dis,S}+\ev{S}_{S}.
\end{split} 
\end{equation}

\begin{figure}[h]
\centerline{
\includegraphics[width=9 cm]{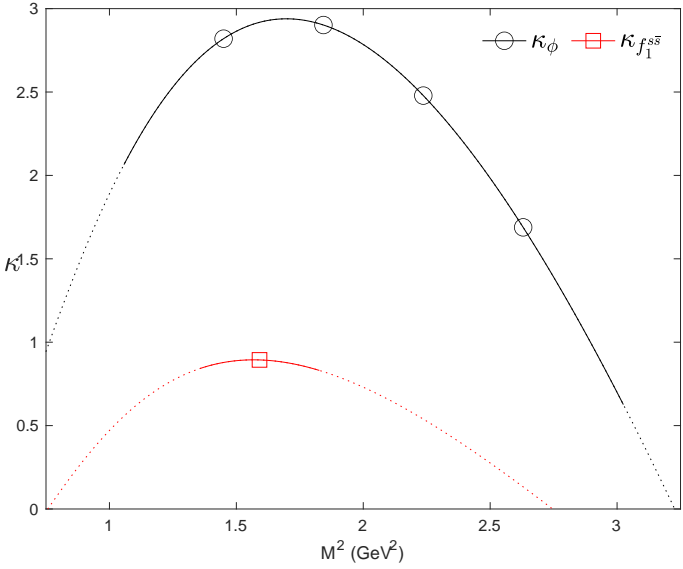}}
\caption{Borel curves for the auxiliary parameters $\kappa_{\phi}$ (black line) and $\kappa_{f_1^{s\bar{s}}}$ (red line), obtained from our vacuum sum rule analysis at the renormalization scale of 2 GeV. The extent of the solid lines indicate the range of the Borel window for each channel.}
\label{kappa_phi}
\end{figure}

\begin{figure}[h]
\centerline{
\includegraphics[width=9 cm]{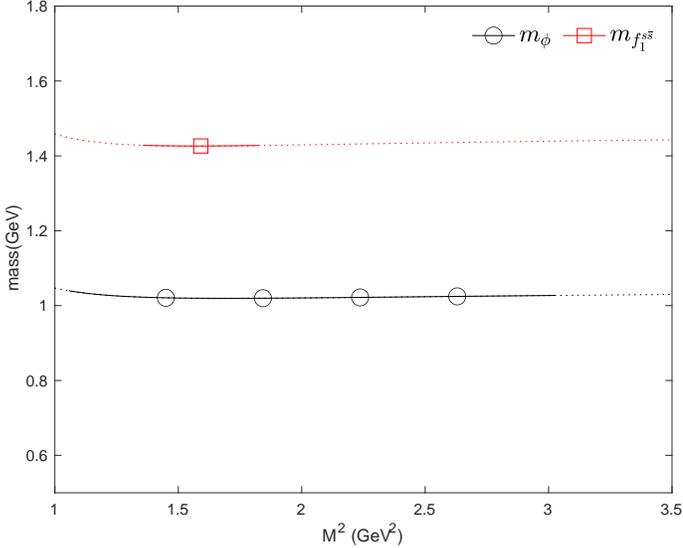}}
\caption{Borel curves for the mass of the $\phi$ (black line) and $f_1^{s\bar{s}}$ (red line) mesons in vacuum. The extent of the solid lines indicate the range of the Borel window for each channel.}
\label{mass_phi}
\end{figure}

To evaluate the chiral breaking and symmetric contributions, we assume that the disconnected condensate values do not depend on their flavour. The light-flavour disconnected four-quark condensates appearing in the sum rules of the $\omega$ and $f_{1}^{s\bar{s}}$ channels, were evaluated in Ref.~\cite{Kim:2021xyp}. Taking into account that there are two light flavors compared to one strange, we get
\begin{equation} 
\begin{split} 
\ev{(\bar{s}\gm \gm[5] \ld s)^{2}}_{dis,S} =&\; \frac{1}{4} \ev{(\bar{q} \gm \gm[5] \ld q)^{2}}_{dis,S},\\
\ev{(\bar{s}\gm  \ld s)^{2}}_{dis,S} =&\; \frac{1}{4} \ev{(\bar{q} \gm  \ld q)^{2}}_{dis,S},
\end{split} 
\end{equation}
where $q$ stands for the isospin doublet. The values of the light flavor four-quark condensates can be obtained from the following equations, discussed in detail in Ref.~\cite{Kim:2021xyp}.
\begin{equation} 
\begin{split} 
(\kappa_{\omega} - \kappa_{\rho}) \frac{448}{81}\ev{\bar{u}u}^{2} =&\; 2\ev{(\bar{q} \gm \gm[5] \ld q)^{2}}_{dis,S},\\
-(\kappa_{f_{1}^{q\bar{q}}} - \kappa_{a_{1}}) \frac{704}{81}\ev{\bar{u}u}^{2} =&\; 2\ev{(\bar{q} \gm \ld q)^{2}}_{dis,S}.
\end{split}  \label{add-discon}
\end{equation}
In the second line, $f^{q\bar{q}}_{1}$ stands for $f_{1}^{s\bar{s}}(1285)$, the parity partner of the $\omega$ meson. The two disconnected four-quark operators on the right hand side of Eq.~\ref{add-discon} are chiral symmetric operators and are, respectively, the additional four-quark operators that appear in the $(\omega,f_1^{q\bar{q}})$ sum rules. While these operators are responsible for the difference between the isospin singlet and triplet currents, and spoil the chiral parity relation between $(\omega,f_1^{q\bar{q}})$, their values can be obtained by combining the results from $(\rho, a_1)$.
The numerical results of the analysis performed in Ref.~\cite{Kim:2021xyp} are given in the first four lines of Table\,\ref{vacuum_result}.
The 5th and 6th lines show the results for the $\kappa$ values in the $\phi$ and $f_1^{s\bar{s}}$ channels obtained from Eq.~\eqref{phen_der} with phenomenological inputs for the corresponding masses and widths. Fig. \ref{kappa_phi} depicts the Borel curves  for  $\kappa_\phi$ and $\kappa_{f_1^{s\bar{s}}}$, while Fig. \ref{mass_phi} shows the Borel curve for the corresponding masses, using the obtained $\kappa$ values demonstrating a stable Borel curve and thus confirming our analysis.

\begin{center}
\begingroup
\setlength{\tabcolsep}{8.5pt} 
\renewcommand{\arraystretch}{1} 
\begin{table}[htbp]
\begin{tabular}{c| c c} \hline
$\mu$ & 1 GeV  & 2 GeV \\\hline
$\kappa_{\rho}(\sqrt{s_0})$     &2.0532(1.223) & 1.6859(1.224) \\
$\kappa_{a_{1}}(\sqrt{s_0})$    &0.4481(1.594) & 0.3637(1.593) \\
$\kappa_{\omega}(\sqrt{s_0})$   &2.4104(1.203) & 1.9884(1.203) \\
$\kappa_{f_{1}^{q\bar{q}}}(\sqrt{s_0})$&1.3204(1.621) & 1.0756(1.621) \\
$\kappa_{\phi}(\sqrt{s_0})$& 2.0056(1.438) & 2.9390(1.448) \\
$\kappa_{f_{1}^{s\bar{s}}}(\sqrt{s_0})$& 0.5495(1.770) & 0.8943(1.772)   \\ 
$\alpha_{s}\ev{D_{V}}_{dis,S}$ &$-(0.2738 \;\mathrm{GeV})^{6}$ & $-(0.2701 \;\mathrm{GeV})^6$\\
$\alpha_{s}\ev{D_{A}}_{dis,S}$ &$(0.2188\;\mathrm{GeV})^{6}$ &$(0.2172 \;\mathrm{GeV})^6 $\\
$\alpha_{s}\ev{B_{ss}}_{B}$ &(0.2060 GeV)$^6$ & (0.2375 GeV)$^6$\\
\hline 
\end{tabular}
\caption{The values of the $\kappa$ parameters and four-quark condensates evaluated from our vacuum analysis, given in units of GeV. $D_{V}$ and $D_{A}$ stand for $(\bar{q}\gm \ld q)^{2}$ and $(\bar{q}\gm \gm[5] \ld q)^{2}$, respectively.}
\label{vacuum_result}
\end{table}
\end{center}
\endgroup


\subsection{Nuclear Matter Analysis \label{sec:nuclear_matter}}
After having fixed the $\kappa$ parameters in vacuum, we can now proceed to the sum rule analysis at finite density. The modifications of the strange quark and gluon condensates at leading order in density $\rho$ can be given as \cite{Cohen:1991nk}
\begin{equation} 
\begin{split} 
\ev{\bar{s}s}_{\rho} \simeq&\; \ev{\bar{s}s} + \ev{\bar{s}s}{N} \rho = \ev{\bar{s}s} + \frac{\sigma_{sN}}{m_{s}} \rho, \label{sbars_density} \\
\ev{\frac{\als}{\pi}G^{2}}_{\rho} \simeq&\; \ev{\frac{\als}{\pi}G^{2}} + \ev{\frac{\als}{\pi}G^{2}}{N}\rho\\
=&\; \ev{\frac{\als}{\pi}G^{2}} - \frac{8}{9}(M_{N} - \sigma_{\pi N} -\sigma_{sN}) \rho,
\end{split} 
\end{equation}
where $M_{N}$ is the nucleon mass, $\sigma_{\pi N} = 2m_{q} \ev{\bar{q}q}{N}$ the $\pi N$ sigma term and $\sigma_{sN} = m_{s} \ev{\bar{s}s}{N}$ the strangeness sigma term. Including the twist-2 and -4 operators, the coefficients of Eq.~(\ref{eq:OPE_coeff}) are at finite density modified as
\begin{equation} 
\begin{split} 
c_{0}(\rho) =&\; c_{0}(0),\\
c_{2}(\rho) =&\; c_{2}(0),\\
c_{4}(\rho) =&\; c_{4}(0) + \rho\bigg[-\frac{2}{27}M_{N}+\frac{56}{27}\sigma_{sN}\\
&\;+\frac{2}{27}\sigma_{\pi N} + A_{2}^{s}M_{N} -\frac{7}{12}\frac{\als}{\pi} A^{g}_{2}M_{N}\bigg]\\
c_{6}(\rho) =&\; c_{6}' + \rho \bigg[-2XM_{N}^{2}-\frac{104}{81}m_{s}^{2}\sigma_{sN} + \frac{4}{81}m_{s}^{2}\sigma_{\pi N} 
\\&\;-\frac{3}{4} m_{s}^{2} A_{2}^{s} M_{N} -\frac{5}{6}A_{4}^{s} M^{3}_{N}\bigg] ,
\end{split} 
\end{equation}
where 
\begin{equation} 
\begin{split} 
c'_{6} =&\; -\pi \als \bigg(\frac{7}{9}\ev{B_{ss}}_{B}\frac{\ev{\bar{s}s}_{\rho}^{2}}{\ev{\bar{s}s}^{2}} + \ev{S^{\phi}}_{S} \bigg)\\
&+\frac{m^{2}_{s}}{6}\bigg[\frac{1}{3}\glucon-8m_{s}\ev{\bar{s}s}  \bigg],\\
\ev{S^{\phi}}_S =&\;\ev{(\bar{s}\gm \gm[5] \ld s)^{2}}_{dis,S} + \ev{S}_{S}.
\end{split} 
\end{equation}

Let us here briefly outline our treatment of the density dependence of the four-quark condensates appearing in the 
$\phi$ meson sum rule. The chiral symmetry breaking operator $\ev{B_{ss}}_{B}$ is assumed to take a form proportional to 
$\ev{\bar{s}s}_{\rho}^2$, in which $\ev{\bar{s}s}_{\rho}$ is modified according to Eq.\,(\ref{sbars_density}). The behavior of the chiral symmetric operators $\ev{S}_{S}$ 
at finite density is much less clear, because the restoration of 
chiral symmetry cannot be used here as a guiding principle. 
To explore the full range of possibilities, we 
consider two extreme cases, which will then be used to estimate 
the uncertainties of the final results caused by this unknown 
density dependence of $\ev{S}_{S}$. 
First, we assume the terms in $\ev{S}_{S}$ to remain constant at their vacuum value, without any 
density dependence. For the second case, we take them to have the same 
relative modification as the chiral symmetry breaking term explained above.

The results of our analysis are shown in Fig.\,\ref{kappa_phi}, together 
with the result reported by the KEK E325 
experiment \cite{KEK-PS-E325:2005wbm} (red square).
\begin{figure}[h]
\centerline{
\includegraphics[width=9 cm]{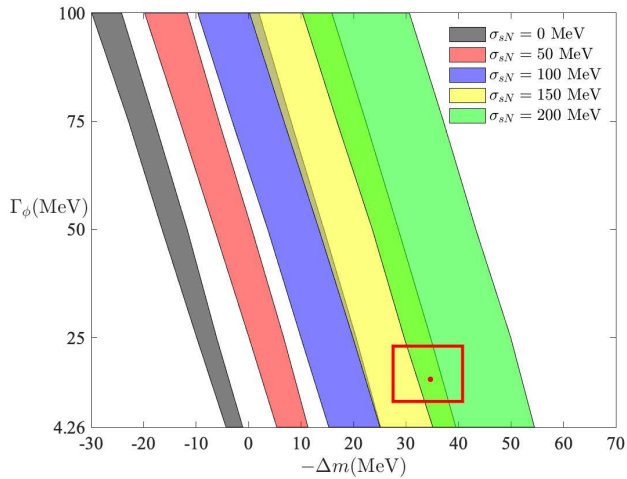}}
\caption{The negative mass shift of $\phi$ meson versus its decay width using the value of disconnected four-quark condensates from the light meson cases. 
The red square indicates the experimental result of the KEK E325 Collaboration \cite{KEK-PS-E325:2005wbm}. The definition of the left and right boundaries are explained in the text.}
\label{equi_pot}
\end{figure}

The boundaries of the color bands in this plot are obtained from two principle sources of error. 
The first is the uncertainty coming from the choice of the renormalization scale ($\mu = 1$ GeV or $\mu = 2$ GeV), 
which dominates the error for small 
$\sigma_{sN}$ values. The second is related to the treatment of the 
chiral symmetric four-quark condensates. As discussed above, we consider two 
extreme cases and estimate the uncertainty introduced by this choice. 
The later error increases with increasing $\sigma_{sN}$, and hence leads to 
thicker error bands for larger $\sigma_{sN}$ values. 
As can be seen in 
Fig.\,\ref{equi_pot}, the latter error dominates for $\sigma_{sN} \gtrsim 100$ MeV.

Finally, inspecting Fig.\,\ref{equi_pot}, one notes that the solutions that satisfy the sum rules 
with different mass shift and decay width values lie on an essentially linear curve, which allows us to 
express our result in a simple analytic expression.
By fitting the curves of Fig.\,\ref{equi_pot}, we obtain
\begin{equation} 
\begin{split} 
\Gamma_{\phi} = \; a\Delta m_{\phi} + b\sigma_{sN}+c,
\end{split} 
\label{fit_result}
\end{equation}
where $a = (3.947^{+0.139}_{-0.130})$, $b = (0.936^{+0.180}_{-0.177} )$ and $c = -(7.707^{+4.791}_{-5.679}) \mathrm{MeV}$.

\section{Discussion \label{sec:discussion}}
The main result of our in-medium analysis is given in Fig.\,\ref{equi_pot} and 
Eq.\,(\ref{fit_result}). 
For a fixed value of $\sigma_{sN}$, we see that the sum rules allow for either a large increase of the decay 
width or (for most cases) a mass reduction for the $\phi$ meson in nuclear matter (or a combination of the two). 
Without any additional input, it is not possible to distinguish the different scenarios. However, limiting the 
in medium decay width to values below 50 MeV, which lies within predictions of most effective model calculations 
\cite{Gubler:2015yna,Gubler:2016itj,Cabrera:2016rnc,Cabrera:2017agk,Cobos-Martinez:2017vtr,Cobos-Martinez:2017woo}, 
one can determine the $\phi$ meson nuclear matter mass shift with a precision of about 10 MeV. 

\begin{figure}[h]
\centerline{
\includegraphics[width=9 cm]{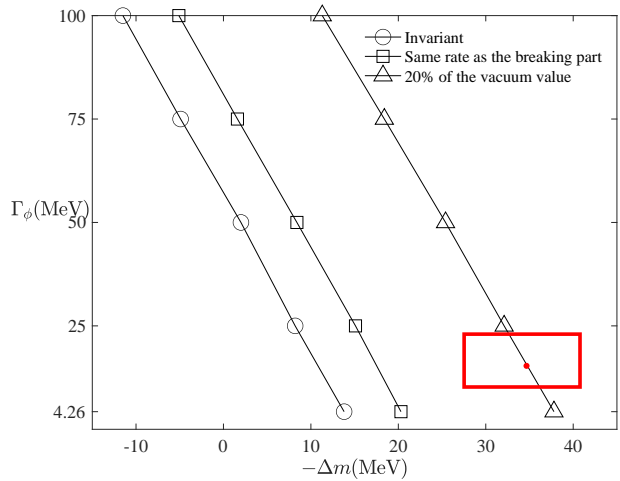}}
\caption{The negative mass shift of the $\phi$ meson versus its decay width with the input parameter values renormalized at 1 GeV, $\sigma_{sN}=$ 50 MeV and the symmetric four-quark condensates, kept invariant (circle), changed in the same rate as the breaking part (square) and reduced to 20\% of the vacuum value (triangle).
The red square indicates the experimental result of the KEK E325 Collaboration \cite{KEK-PS-E325:2005wbm}}
\label{diff_sym}
\end{figure}

Conversely, using the available experimental results on the $\phi$ meson decay width and mass shift in 
nuclear matter, the result of Fig.\,\ref{equi_pot} allows us to determine a range of $\sigma_{sN}$ values that 
are consistent with the sum rules. The finding of the KEK E325 experiment \cite{KEK-PS-E325:2005wbm}, 
indicated as a red square in Fig.\,\ref{equi_pot}, can for instance understood to be consistent with a strangeness sigma term of about $\sigma_{sN} \sim 125$ MeV. As previously pointed out in Ref.\,\cite{Gubler:2014pta} and again confirmed here, this finding does not agree with recent lattice QCD calculations of this quantity \cite{Alexandrou:2019brg,Borsanyi:2020bpd,Copeland:2021qni}, which get values of around $\sigma_{sN} \sim 50$ MeV (see also Refs.\cite{Gubler:2018ctz,Aoki:2021kgd} for compilations and detailed discussions of the lattice QCD calculations). It will be interesting to see whether the J-PARC E16 experiment and further experimental studies at HADES and ALICE will confirm the KEK E325 result and if the above-mentioned inconsistency persists. It is also possible that the discrepancy can be resolved by allowing a much larger change in the chiral symmetric four-quark condensate. In fact, as can be seen in Fig. \ref{diff_sym} allowing the chiral symmetric operator to decrease to about 20\% (45\%) of the vacuum value with parameters chosen at $\mu=1$ (2) GeV, one finds that the KEK E325 result could be consistent with $\sigma_{sN} \sim 50$ MeV. We would then have to understand the origin of the large medium effect of the chiral symmetric operators.

\section{Conclusion and Outlook \label{sec:conclusion}}
We have in this work studied the behavior of the $\phi$ meson in nuclear matter, making use of a QCD sum rule approach that takes the chiral properties of the four-quark condensates properly into account and furthermore allows for not only an in-medium $\phi$ meson mass shift, but also broadening in the phenomenological parametrization of the spectral function. 
As a result, we find that while the sum rules can be satisfied by multiple combinations of mass shift and broadening scenarios, the range of allowed modification parameters can be strongly constrained by the value of the strangeness sigma term $\sigma_{sN}$. Specifically, an increasing value of $\sigma_{sN}$ will lead to a larger negative mass shift, but not much influence the magnitude of the in-medium decay width (see Fig.\,\ref{equi_pot}).

In the future, two major items could be considered to further improve the accuracy of the QCD sum rule predictions given in this work. First, non-zero momentum effects should be taken into account along the lines of Refs.\cite{Lee:1997zta,Kim:2019ybi}, especially in view of the fact that most 
experimentally measured $\phi$ mesons in nuclear matter are not at rest with respect to the medium and finite momentum effects therefore should be accounted for. Second, the finite-density behavior of the chiral symmetric four-quark condensates are still largely unknown, the related uncertainty hence substantially contributing to the size of the error bands shown in Fig.\,\ref{equi_pot}. It remains a challenging theoretical task to constrain this behavior through model or lattice QCD calculations. 

On the experimental side, we can expect a lot of new data coming out during the next few years, not only from the ongoing J-PARC E16 experiment \cite{Ashikaga:2019jpc}, but also more precise data becoming available from experiments at HADES \cite{HADES:2018qkj} and ALICE \cite{ALICE:2021cpv}. We expect that these new data, combined with further improved theoretical calculations, will lead to a more complete understanding of the behavior of the $\phi$ meson in nuclear matter and, eventually, of the relationship between its properties and chiral symmetry of QCD.

\section*{Acknowledgements}
J.K. and S.H.L. were supported by Samsung Science and Technology Foundation under Project Number SSTF-BA1901-04. 
P.G. is supported by the Grant-inAid for Scientific Research (C) (JSPS KAKENHI Grant Number JP20K03940) and the Leading Initiative for Excellent Young Researchers (LEADER) of the Japan Society for the Promotion of Science (JSPS).


\end{document}